\documentclass[12pt]{iopart}

\usepackage{graphicx}
\usepackage{amssymb}
\begin{document}

\title[Joint neutrino and gravitational-wave search for nearby supernovae]
{Searching for prompt signatures of nearby core-collapse supernovae by a 
joint analysis of neutrino and gravitational-wave data}

\author
{I Leonor$^1$, L Cadonati$^2$, E Coccia$^{3,4}$,
S D'Antonio$^4$,\\ A Di Credico$^3$, V Fafone$^4$,
R Frey$^1$, W Fulgione$^5$,\\ E Katsavounidis$^6$,
C D Ott$^7$, G Pagliaroli$^3$, K Scholberg$^8$,\\
E Thrane$^9$, F Vissani$^3$}

\address{$^1$ University of Oregon, Eugene, OR 97403, USA}
\address{$^2$ University of Massachusetts, Amherst, MA 01003, USA}
\address{$^3$ INFN Laboratori Nazionali del Gran Sasso, Assergi, Italy}
\address{$^4$ Universit\`{a} di Roma Tor Vergata, Roma, Italy}
\address{$^5$ INAF, Torino, Italy}
\address{$^6$ Massachusetts Institute of Technology, Cambridge, MA 02139, USA}
\address{$^7$ California Institute of Technology, Pasadena, CA 91125, USA}
\address{$^8$ Duke University, Durham, NC 27708, USA}
\address{$^9$ University of Minnessota, Minneapolis, MN 55455, USA}

\ead{ileonor@uoregon.edu}

\begin{abstract}
We discuss the science motivations and prospects for a joint analysis of
gravitational-wave (GW) and low-energy neutrino data to search for prompt
signals from nearby supernovae (SNe).  Both gravitational-wave and low-energy
neutrinos are expected to be produced in the innermost region of a core-collapse
supernova, and a search for coincident signals would probe the processes
which power a supernova explosion.  It is estimated that the current generation 
of neutrino and gravitational-wave detectors would be sensitive to Galactic 
core-collapse supernovae, and would also be able to detect electromagnetically 
dark SNe.  A joint GW-neutrino search would enable improvements to searches 
by way of lower detection thresholds, larger distance range, better live-time 
coverage by a network of GW and neutrino detectors, and increased significance 
of candidate detections.  A close collaboration between the GW and neutrino 
communities for such a search will thus go far toward realizing a much 
sought-after astrophysics goal of detecting the next nearby supernova.
\end{abstract}

\submitto{\CQG}
\maketitle

\section{Motivation}
The predicted rate of core-collapse supernovae (SNe) in our Galaxy is $\sim2$
per century; the rate is about twice the Galactic rate out to the Andromeda
galaxy ($\sim1$~Mpc), and about one per year out to the Virgo cluster
($\sim10$~Mpc) \cite{ando05}.  These numbers are accompanied by large
uncertainties inherent in converting galaxy properties to supernova rates.
However, \cite{ando05} also suggests that the increased number of nearby
core-collapse supernovae discovered within the past few years strongly
indicates that the predicted rates might be significantly underestimated by a
factor of $\sim3$ in the $3-5$~Mpc range.  A direct upper limit on the Galactic
SNe rate, based on non-observations of antineutrino events in the past 25
years, is given in \cite{strumia08}.

Contemporary neutrino detectors and kilometer-scale gravitational-wave
(GW) detectors are currently poised to detect the next Galactic
core-collapse supernova.  Both neutrino and gravitational-wave signals
are expected to be generated in the innermost region of a dying star,
and both signals are expected to be emitted within a short time
interval of each other, i.e.  within a few milliseconds
\cite{ott09,ott09_2}.  While both neutrino and GW detectors are
preparing to independently detect a Galactic supernova, there are
science benefits to systematically searching for a supernova signature
using a joint analysis of neutrino and GW data which is guided by the
expected proximity of the neutrino and GW signals.  These science
benefits include lower detection threshold requirements, better
live-time coverage, increased significance of candidate detections,
extended distance reach to the local volume of galaxies, and increased
sensitivity to core-collapse events which have only a weak or
non-existent electromagnetic signature.

\begin{figure}
\begin{center}
\includegraphics[height=3.5in]{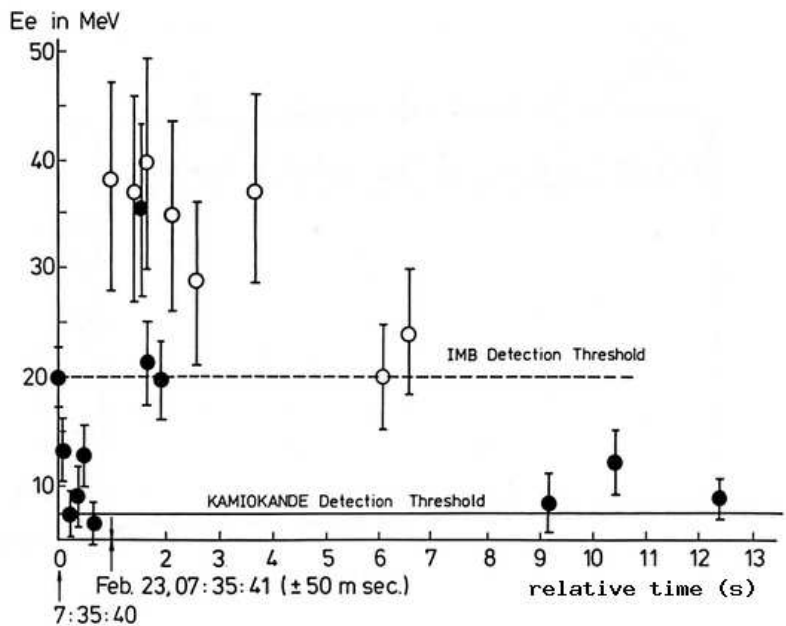}
\end{center}
\caption{\label{fig:sn1987a}The energy of the events detected by 
Kamiokande II (closed circles) and IMB (open circles), produced by the neutrino 
burst from SN1987A, as a function of time, in seconds \cite{koshiba88}.}
\end{figure}

The detection of a burst of low-energy neutrinos from SN1987A, at a
distance of about 50~kpc in the Large Magellanic Cloud, by the
Kamiokande II and Irvine-Michigan-Brookhaven (IMB) experiments, and by
scintillation neutrino detectors
\cite{hirata87,bionta87,koshiba88,alekseev87,aglietta87} demonstrated
the capability of neutrino detectors to detect SN events, and paved
the way for a description and validation of the standard model of
neutrino emission from a core-collapse supernova.  In a core-collapse
supernova, $\sim99\%$ of the neutron star's gravitational binding
energy ($\sim3\times10^{53}$~ergs) is released in the form of
neutrinos (and anti-neutrinos) of all flavors.  These neutrinos have
energies in the few tens of MeV and are emitted over a time scale of a
few tens of seconds.  The neutrino light curve is expected to show
structure, with an increase in luminosity during the first
$\sim0.5$~second due to accretion of matter onto the proto-neutron
star \cite{totani98,loredo02,pagliaroli09_1}.  A neutronization burst, which is
a peak in the $\nu_e$ luminosity produced as the shock from the core
bounce propagates through the star's outer core, is expected to last a
few milliseconds after the core bounce.  However, the energy emitted
in the neutronization burst is only $\lesssim1\%$ of the total energy
emitted, and such a feature might not be easily recognized in a
measured neutrino light curve because of its short duration, and
because the cross section of $\nu_ee$ scattering is lower than that of
the dominant inverse beta decay ($\bar{\nu}_ep\rightarrow e^+n$)
reaction in a neutrino detector's medium \cite{totani98}.  The events
detected by Kamiokande II and IMB from SN1987A are reproduced in
figure \ref{fig:sn1987a}.

The groundbreaking discovery of a neutrino burst from SN1987A was
guided by an optical sighting of the supernova
\cite{hirata87,koshiba88}.  The optical sighting of such a close
astrophysical event motivated the analysis of archived neutrino data
and guided the time scale in which to perform the search.  It is
plausible, however, that a fraction of core-collapse events are
accompanied only by a faint electromagnetic display.  This might be
due to extinction brought about by an extremely dusty environment or
the intervening interstellar medium, or an inherently weak
accompanying electromagnetic emission with a fast decay time
\cite{kochanek08}.  For example, the supernova of Cassiopeia~A, one of
the youngest known Galactic supernova remnants and which is at a
relatively close distance of 3.4~kpc, has no historical record of
widespread sighting \cite{dasilva93,hughes80} during the epoch when
the explosive fireworks would have reached Earth around 340 years ago,
and it is plausible that this supernova fell in this category
\cite{dasilva93,krause08,young06}.  Obscuration by dust could also
explain the non-sighting of the supernova of an even younger,
$\lesssim150$ year-old remnant, G1.9+0.3 \cite{green08}.  This would
be analogous to how the dearth of known Galactic supernova remnants
could possibly be attributed to low surface brightness, leaving faint
supernova remnants unresolved from the Galactic background emission
\cite{green05}.

It has also been argued that observations suggest a deficit of optically 
observed high-mass ($\gtrsim25{\rm M}_{\odot}$) core-collapse SN progenitors 
\cite{kochanek08}, and that optical searches have provided little information on 
the possibility that massive stars end their lives by forming black holes 
without the dramatic electromagnetic signature of an explosion.  Reference 
\cite{kochanek08} estimates that an upper bound on the rate of so-called 
``failed" supernovae is roughly equal to the rate of successful SNe, and that 
the lower bound on black hole formation rate is $\sim25\%$ that of normal SNe.

A joint GW-neutrino search for nearby core-collapse supernovae could potentially
provide insight to this scenario where a considerable fraction of stars end
their lives with little or no electromagnetic display.

Indeed, the natural progression of the expected supernova signature---from the 
prompt GW and neutrino signals to the optical signal which is expected to rise
many minutes to hours later---and the current state of neutrino and 
gravitational-wave detectors and their respective detection algorithms, make it 
likely that the detection of the next nearby supernova will proceed in a 
direction opposite that of SN1987A, i.e. that prompt neutrino and GW detections 
would trigger optical telescopes to search for an optical counterpart.  The 
infrastracture of the Supernova Early Warning System (SNEWS), for example, is 
designed to alert observatories in the event of a detection of neutrinos from a 
nearby supernova \cite{schol04,schol08}.  Simulations also indicate that, for a 
Galactic supernova, the neutrino detectors Super-Kamiokande and IceCube would be 
able to reconstruct the SN bounce time to within a few milliseconds 
\cite{pagliaroli09,halzen09}.  On the gravitational-wave side, there is also an 
alert system, called LOOC UP \cite{kanner08}, which is being developed to send 
plausible future candidate GW triggers to optical observatories for confirmation 
of a corresponding astrophysical source.

Astrophysical events---such as gamma-ray bursts (GRBs) and flares from soft 
gamma repeaters (SGRs)---detected by other observatories have been extensively 
utilized as external triggers in the analysis of LIGO-Virgo data to search for 
GW counterparts to these events 
\cite{grb_s5,grb070201_07,multigrb07,abbottgrb05,sgrstorm09,sgr08}.  The
strategy of using external triggers with precise event timing and position 
information to look for GW signals is motivated mainly by the decrease in both
background rate and the effective number of experimental trials that shorter 
analysis time windows make possible.  In the case of GRBs and SGRs, the analysis 
windows range from a few seconds to a few minutes.  On the other hand, the use 
of an optical signal from a supernova as an external trigger does not provide 
the same tight constraints on the time and duration of the analysis window.  
Studies indicate that, at best, the time of a supernova explosion can be 
determined from an optical light curve to within a few hours, but only if the 
first measurement of the optical flux is made within a day of the explosion 
\cite{cowen09}.

In contrast, the gravitational-wave and neutrino signals are expected to be 
detected within a tight window, ranging from a few milliseconds to a few 
seconds, depending on the dominant GW emission process \cite{ott09,ott09_2}.

\section{Science benefits of a joint GW-neutrino search}

Several of the world's neutrino detectors have performed searches for
core-collapse supernovae and have evaluated their respective detection 
efficiencies as a function of distance.  The Super-Kamiokande (Super-K) water
Cherenkov detector in Japan \cite{ikeda07}, the scintillation detectors Large 
Volume Detector (LVD) \cite{agafonava08} and Borexino \cite{cadonati02} in 
Italy, and KamLAND in Japan \cite{piepke01}, IceCube at the South Pole 
\cite{kowarik09}, MiniBooNE in the USA \cite{aa09}, and others 
\cite{schol08_2} are expected to robustly detect a neutrino burst 
from a Galactic supernova in the baseline model scenario.  Super-K, for example, 
would detect $\sim8000$ events for a core-collapse SN at the center of the 
Milky Way, $\sim8.5$~kpc away \cite{ikeda07}.  The Baksan scintillation detector 
had previously also performed a systematic search \cite{alexeyev02}.

Analogously, all-sky searches for GW bursts have been performed using data from
the Laser Interferometer Gravitational-Wave Observatory (LIGO), the most recent
of which made use of data from the fourth (S4) and fifth (S5) LIGO science runs 
\cite{bursts4,bursts5}.  While no gravitational waves have been directly 
detected from an astrophysical source, the current generation of LIGO-Virgo 
interferometric GW detectors have made tremendous progress in improving their 
sensitivities \cite{ligo09,smith09,abbott04,virgo,geo}, and are expected to be sensitive to 
several models of GW emission from a Galactic core-collapse supernova 
\cite{ott09,ott09_2}.  Significant improvements in sensitivy are expected to
continue with the anticipated advent of the next generation of GW interferometers,
Advancd LIGO and Virgo.  However, while the astrophysical motivation for expecting 
gravitational waves to accompany core-collapse supernovae is strong, the 
expected rate, gravitational-wave strength and waveform morphology are uncertain
\cite{ott09,ott09_2}.  As a benchmark, the expected energy going into 
gravitational waves may range from 
$10^{-10}$ to $10^{-4}$~M$_\odot\,c^2$ (or $2\times 10^{44}$ to $2\times 10^{50}$~ergs),
and thus only a small fraction of the energy liberated in neutrinos.

Estimating the sensitivity improvement of a gravitational-wave search
due to a tighter search window and lower background rate requires
assumptions on the spectrum of the background events.  In the all-sky
GW burst search using the first-year data of LIGO's fifth science run,
a false alarm rate of 1 in 100 years in the 64-200~Hz frequency band
corresponds to a three-detector network signal-to-noise ratio (SNR)
threshold of $\approx 8.5$.  For frequencies above 200~Hz, the
corresponding SNR threshold is lower, but the interferometers' strain
sensitivity is lower at this frequency band (cf. figure~5 and figure~2 
of \cite{bursts5}, and also Appendix~E of \cite{bursts5}).  Requiring 
a coincidence of GW events within O(1s) of a neutrino signal tuned at a 
rate of 1 per day would allow GW detectors to operate at a false alarm 
rate of $3\times 10^{-5}~$Hz, which in turn corresponds to a SNR 
threshold of $\approx 3.5$~\cite{bursts5}, or an improvement of a factor 
of $\sim2$ in sensitivity.  The distance reach of gravitational-wave 
detectors scales linearly with the inverse of SNR.  Such potential 
improvement in gravitational-wave sensitivity, in a joint GW-neutrino 
search, will increase the science reach of the GW instruments relative 
to what they can achieve alone.

To quantify the science reach, it is important to appreciate that the
guidance from source phenomenology is subject to significant
uncertainties \cite{ott09_2}.  All estimates of GW bursts associated
with supernovae rely on models (see
\cite{ott:06prl,dimmelmeier:08,abdikamalov:10,
scheidegger:08,marek:09,murphy:09,kotake:09}
for recent GW emission estimates). Most such models are not yet
3-dimensional, do not incorporate the entire set of possibly relevant
physics, and, most importantly, do not (in most cases) predict robust
supernova explosions as observed in the electromagnetic universe.
Moreover, none of the current ``state-of-the-art'' simulations can
make reliable predictions of the mechanism responsible for the
observed velocities of pulsars of up to 1000 km/s (but see \cite{scheck:06}
who do predict such kicks, albeit with a simplified model).  It is likely that
these velocities were imparted on the neutron stars at birth (``pulsar
birth kicks'') which obviously must involve multi-dimensional dynamics
and gravitational-wave emission presently not accounted for in models.
Thus, despite the availability of multiple potential explosion
mechanisms and their associated multi-dimensional dynamics and
gravitational-wave signatures, the current picture is unlikely to be
complete.  Moreover, the current, most pessimistic estimates are
probably overly so in predicting the gravitational-wave yield of
core-collapse supernovae.  In the absence of complete models,
observations can and must guide our understanding of the astrophysical
systems.  Joint analysis of neutrino data with initial and enhanced
LIGO-Virgo observations would significantly enhance the capability to
constrain models of gravitational-wave emission in core-collapse
events.

\begin{figure}
\begin{center}
\includegraphics[width=5.0in]{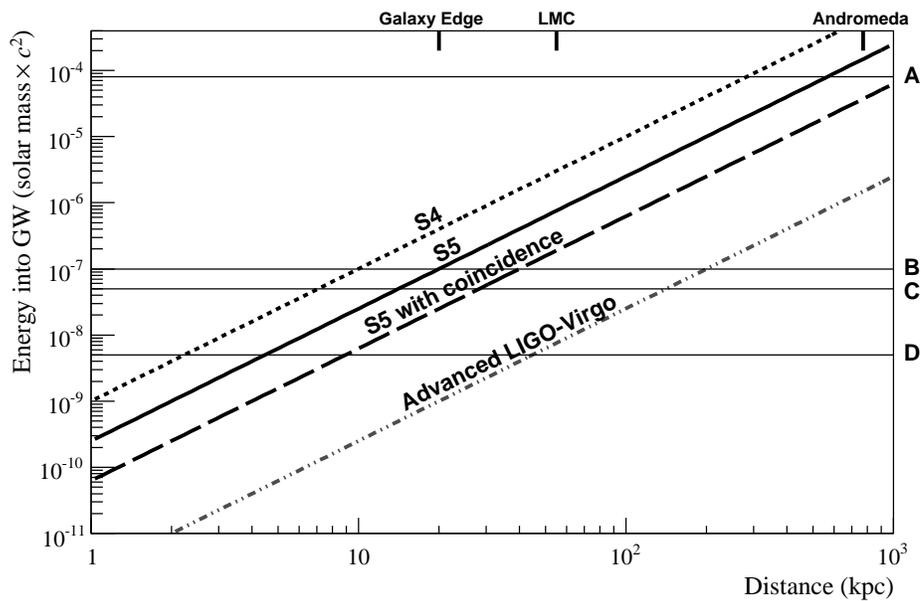}
\end{center}
\caption{\label{fig:gwsens}Sensitivity of gravitational-wave detectors to 
core-collapse supernovae as limited by the efficiency of past, present and 
foreseen searches.  The diagonal lines in this log-log plot reflect the 
fundamental relation connecting energy going into gravitational waves, distance 
to the source, and root-sum-square strain sensitivity of a search, $h_{\rm rss}$
(cf. equation~\ref{eq:energy}).  These lines correspond to a fixed search 
sensitivity {\it at} the gravitational-wave detectors for narrow-band signals 
in the most sensitive frequency region ($\sim150$~Hz).  All combinations of 
energy-distance above and to the left of these diagonal lines can be probed in a 
search.  The dotted line identified as ``S4'' corresponds to LIGO's fourth 
science run, and the solid line labelled ``S5'' to LIGO's fifth science run.  
The dashed line reflects the estimated improvement in sensitivity in a joint 
GW-neutrino search.  The dot-dashed line is the expected reach of Advanced LIGO 
and Virgo, with no assumption of joint searches with neutrinos \cite{smith09}.  
The horizontal lines represent upper bounds on the energy release for four core 
collapse models (as summarized in \cite{ott09_2}):
           A: PNS pulsations (acoustic mechanism, \cite{ott:06prl});
           B: rotational instability;
           C: rotating collapse and bounce;
           D: convection and SASI.}
\end{figure}

\begin{figure}
\begin{center}
\includegraphics[width=5.0in]{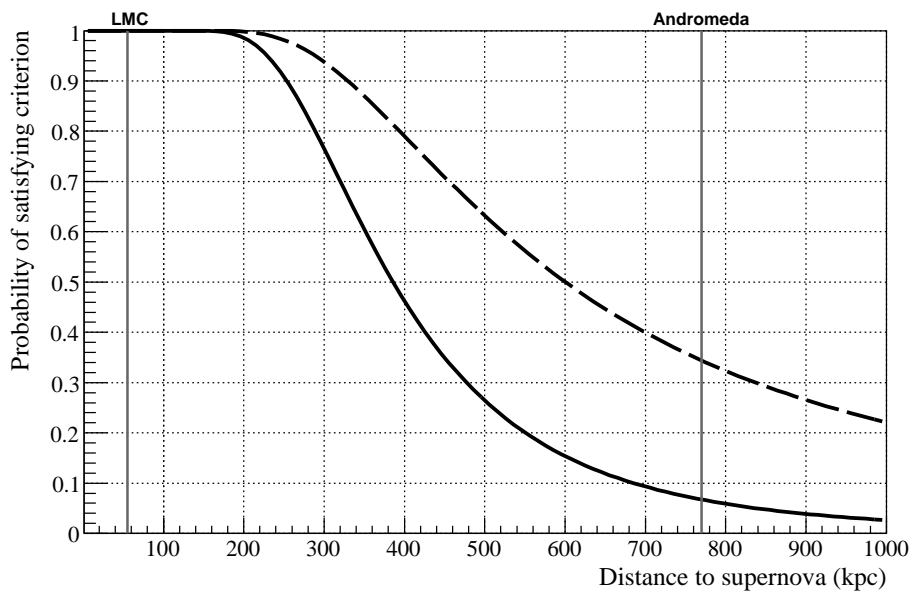}
\end{center}
\caption{\label{fig:nusens}Estimated probability of satisfying a Super-K neutrino 
burst search criterion as a function of distance.  Solid curve: standard search 
parameters \cite{ikeda07}.  Dashed curve: probability if only a single neutrino 
event is required, in coincidence with a GW signal.}
\end{figure}

For isotropic emission of gravitational waves, 
the luminosity distance is related to the energy emitted by the source in
GW waves, $E_{\rm GW}$, and to the root-sum-square strain amplitude at the 
detector, $h_{\rm rss}$, by \cite{bursts4},
\begin{equation}
\label{eq:energy}
E_{\rm GW}\approx \frac{\pi^2 c^3}{G}\,D^2 f_o^2 h_{\rm rss}^2 ~~~,
\end{equation}
\noindent
where $f_{o}$ is the observed frequency of the waves.
For a hypothetical source at a Galactic distance of $10$~kpc and an assumed 
signal morphology of a sine-Gaussian waveform \cite{bursts5} with central 
frequency $153$~Hz and quality factor $Q=9$ (where $Q$ is a dimensionless
quantity which is roughly a measure of the width of the waveform in terms
of number of cycles of the sinusoid), the mass sensitivity during the LIGO 
S5 run is $1.9\times10^{-8}\,\mbox{M}_{\odot}\,c^2$ \cite{bursts5}.  The 
proposed joint search, with a factor of $\sim 2$ improvement in sensitivity, 
would decrease by a factor of $\sim 4$ the minimum energy which is probed in
core-collapse supernovae, i.e. down to $4\times10^{-9}\,\mbox{M}_{\odot}\,c^2$ for 
a Galactic supernova with signal content in the most sensitive band of the 
gravitational-wave detectors.  Searches at higher frequencies would be 
penalized by the $f_{o}^2$ dependence and by lower strain sensitivity of the
detectors at these frequencies.  At the same time, however, such searches are 
characterized by a lower accidental background rate \cite{bursts5}, and the 
improvement in sensitivity which could be achieved at higher frequencies would 
be about the same as that which was estimated for searches at the instruments' 
most sensitive frequency, i.e. a factor $\sim2$ improvement (cf. figure 5 of 
\cite{bursts5}).  These scaling laws, and the potential improvement in science 
reach offered by a joint search, are summarized in figure~\ref{fig:gwsens}, 
together with the expected gravitational-wave emission in four sample emision 
mechanisms.

A gravitational-wave coincidence requirement also has the potential to
improve the sensitivity of neutrino experiments by relaxing the criteria for 
detection.  For example, Super-K's recent ``distant'' burst search 
\cite{ikeda07} requires two neutrino events (with energy threshold 17~MeV) 
within 20 seconds, which corresponds to approximately 8\% probability of 
detecting a supernova in Andromeda.  The accidental coincidence rate for this
criterion is less than one per year; the single event rate at this threshold is 
about 1 per day.  If one could achieve an acceptable accidental rate by 
requiring coincidence of a single neutrino event with a gravitational-wave 
signal, then the probability of a core-collapse event in Andromeda satisfying 
the search criterion would be about 35\%, as shown in figure~\ref{fig:nusens}.  
Distant burst search parameters could be re-optimized with respect to current 
ones; the neutrino event energy threshold could potentially be reduced, further 
improving sensitivity.

\section{Summary}
We have motivated the search for nearby core-collapse supernovae using a joint
analysis of low-energy neutrino and gravitational-wave data, and we have shown 
examples of the science benefits of such a joint analysis.  Turning this idea 
into a reality 
in the immediate future using contemporary neutrino and gravitational-wave data 
would make possible a richer exploration of the innermost, dynamical processes 
in a core-collapse supernova.  A search like this is a necessary complement to
the joint high-energy neutrino and gravitational-wave search that is currently
being planned \cite{gwhen09}.  Moreover, embarking on such a task now would be 
forward-looking, since this kind of analysis would gain importance as the 
sensitivities of experiments improve.  The Advanced LIGO and Virgo detectors
\cite{advLV1} are expected to start operating in 2014, and are designed
to improve on the sensitivity of the initial detector configurations by a factor
of $\sim$10.  At the same time, there are a number of large future neutrino 
experiments planned, employing various technologies, including some of megaton 
scale
\cite{Nakamura:2003hk,Kistler:2008us,MarrodanUndagoitia:2006qs,deBellefon:2006vq,Autiero:2007zj}.
Reference \cite{ando05} points out that such detectors will be able to observe 
on the order of one supernova neutrino event every few years from beyond the 
Local Group of galaxies ($\sim2$~Mpc).  In such a regime, some kind of external 
(non-neutrino) trigger will be essential to distinguish supernova 
neutrino-induced events from background.

\ack
C. D. Ott is partly supported by the National Science Foundation under grant 
number AST-0855535.


\end{document}